\documentclass{emulateapj}

\newcommand{\kms}{\,km\,s$^{-1}$}

\newcommand{\Msun}{\mbox{\,$M_{\odot}$}}
\newcommand{\Lsun}{\mbox{\,$L_{\odot}$}}

\def\spose#1{\hbox to 0pt{#1\hss}}
\def\simlt{\mathrel{\spose{\lower 3pt\hbox{$\mathchar"218$}}
     \raise 2.0pt\hbox{$\mathchar"13C$}}}
\def\simgt{\mathrel{\spose{\lower 3pt\hbox{$\mathchar"218$}}
     \raise 2.0pt\hbox{$\mathchar"13E$}}}

\newcommand{\nmembers}{24}   

\shorttitle{Spectroscopy of Segue\,1}
\shortauthors{Geha~et~al.}

\begin{document}


\title{The Least Luminous Galaxy:  Spectroscopy of the Milky Way Satellite Segue\,1}

%


\author{Marla Geha\altaffilmark{1}}
\altaffiltext{1}{Astronomy Department, Yale University, New Haven, CT~06520.  marla.geha@yale.edu}

\author{Beth Willman\altaffilmark{2}}
\altaffiltext{2}{Haverford College, Department of Physics, 370 Lancaster Avenue, Haverford, PA 19041}

\author{Joshua\ D.\ Simon\altaffilmark{3}}
\altaffiltext{3}{Department of Astronomy, 
       California Institute of Technology, 1200 E. California Blvd.,
       MS 105-24, Pasadena, CA  91125}

     \author{Louis E.\ Strigari\altaffilmark{4,5}}
     \altaffiltext{4}{Kavli Institute for Particle Astrophysics \&
       Cosmology, Physics Department, Stanford University, Stanford,
       CA 94305}
     \altaffiltext{5}{Hubble Fellow}

\author{Evan N.\ Kirby\altaffilmark{6}}
\altaffiltext{6}{UCO/Lick Observatory, University of California,
    Santa Cruz, 1156 High Street, Santa Cruz, CA~95064.}

\author{David R.\ Law\altaffilmark{3}}

\author{Jay Strader\altaffilmark{7}}
\altaffiltext{7}{Harvard-Smithsonian CfA, 60 Garden St., Cambridge, MA 02144}


\begin{abstract}
\renewcommand{\thefootnote}{\fnsymbol{footnote}}

We present Keck/DEIMOS spectroscopy of Segue\,1, an ultra-low
luminosity ($M_V = -1.5^{+0.6}_{-0.8}$) Milky Way satellite
companion. While the combined size and luminosity of Segue\,1 are
consistent with either a globular cluster or a dwarf galaxy, we
present spectroscopic evidence that this object is a dark
matter-dominated dwarf galaxy.  We identify \nmembers\ stars as
members of Segue\,1 with a mean heliocentric recession velocity of
$206 \pm 1.3$\kms. We measure an internal velocity dispersion of
$4.3\pm 1.2$\kms.  Under the assumption that these stars are
gravitationally bound and in dynamical equilibrium, we infer a total
mass of $4.5^{+4.7}_{-2.5} \times 10^5 M_{\odot}$ in the case where
mass-follow-light; using a two-component maximum likelihood model, we
determine a similar mass within the stellar radius of 50\,pc.  This
implies a mass-to-light ratio of ln$(M/L_V) = 7.2^{+1.1}_{-1.2}$ or
$M/L_V = 1320^{+2680}_{-940}$.  The error distribution of the
mass-to-light ratio is nearly log-normal, thus Segue\,1 is dark
matter-dominated at a high significance.  Although Segue\,1 spatially
overlaps the leading arm of the Sagittarius stream, its velocity is
100\kms\ different than that predicted for recent Sagittarius tidal
debris at this position.  We cannot rule out the possibility that
Segue\,1 has been tidally disrupted, but do not find kinematic
evidence supporting tidal effects.  Using spectral synthesis modeling,
we derive a metallicity for the single red giant branch star in our
sample of [Fe/H] $= -3.3\pm0.2$\,dex.  Finally, we discuss the
prospects for detecting gamma-rays from annihilation of dark matter
particles and show that Segue\,1 is the most promising satellite for
indirect dark matter detection.  We conclude that Segue\,1 is the
least luminous of the ultra-faint galaxies recently discovered around
the Milky Way, and is thus the least luminous known galaxy.

\end{abstract}




\section{Introduction}\label{intro_sec}
\renewcommand{\thefootnote}{\fnsymbol{footnote}}

The discovery of ``ultra-faint'' dwarf spheroidal (dSph) galaxies
around the Milky Way has revolutionized our understanding of dwarf
galaxies and their prevalence in the Universe.  These newly discovered
satellites, with total absolute magnitudes fainter than $M_V = -8$,
have all been found in the Sloan Digital Sky Survey (SDSS) via slight
statistical over-densities of individual stars
\citep{willman05b,willman05a,zucker06a,zucker06b,
  belokurov06a,belokurov06b,sakamoto06a, irwin07a, walsh07a}.  These
objects provide important clues to galaxy formation on the smallest
scales \citep{madau08a,ricotti08a} and substantially alleviate the
discrepancy between the observed mass function of Milky Way satellites
and that predicted by standard Lambda Cold Dark Matter models
\citep[][hereafter SG07]{tollerud08a,simon07a}.  \citet{strigari07b} note that the
ultra-faint dSphs have high central dark matter densities and are good
candidates for indirect dark matter detection via gamma-ray emission
by particle annihilation.  Future wide-field surveys that improve on
the sky coverage and photometric depth of the SDSS are likely to
discover many additional ultra-faint Milky Way satellites in the
coming years \citep{koposov07a,walsh08b}.

While the total luminosities of the ultra-faint satellites are
comparable to globular clusters, spectroscopic studies for the
majority of the newly discovered objects firmly suggest that these
objects are dark matter-dominated dwarf galaxies
\citep[][SG07]{kleyna05a,munoz06a,martin07a}. The mass-to-light ratios
for all the ultra-faint dSphs are $M/L_V >
100$~M$_{\odot}$/L$_{\odot}$, with several systems approaching
1000~M$_{\odot}$/L$_{\odot}$, assuming mass-follows-light.
\citet{strigari08a} loosened this constraint, confirming the high
mass-to-light ratios and finding a tight anti-correlation between
mass-to-light ratio and luminosity such that all the Milky Way dwarfs
are consistent with having a common dark matter mass of
$\sim10^7\Msun$ within their central 300\,pc. A theoretical
understanding of the physics that sets the mass-luminosity relation
will provide insight into the formation of galaxies at the smallest
scales.

Further evidence that the ultra-faint satellites are indeed galaxies
comes from metallicity measurements.  The ultra-faint satellites are
the most metal-poor known stellar systems ([Fe/H] $< -2$) and show
internal metallicity spreads up to 0.5~dex in several objects (SG07).
This is in contrast to Milky Way globular clusters which are, on
average, more metal-rich and show little to no internal metallicity
spread \citep[e.g.][]{Pritzl05a}.  In further contrast to globular
clusters, the ultra-faint dwarfs also follow the
luminosity-metallicity relationship established by brighter Milky Way
dwarf galaxies \citep{kirby08b}.  Thus, both the kinematics and
composition of the ultra-faint satellites strongly argue that these
objects are dark matter-dominated galaxies.

The combined size and luminosity of the spectroscopically confirmed
dSphs in the Milky Way are well separated from globular clusters: at a
given luminosity dwarf galaxies have larger sizes and are thus less
compact \citep{belokurov06b,martin08a}.  However, the three faintest
SDSS discoveries, Segue\,1, Willman~1 and Bootes~II, are all in a
region that overlaps with globular clusters.  Studying these extreme
systems should provide important insight to dSphs, and the difference
between dwarfs and star clusters, at all luminosities.  Of these three
objects, only Willman~1 has published kinematics \citep{martin07a}.
Because the systemic velocity of Willman~1 is similar to that of the
foreground Milky Way stars, possible contamination in the kinematic
sample make it difficult to assess whether this object is a dwarf or
globular cluster \citep[][Willman et al.~in prep]{siegel08a}.  Here,
we present the first spectroscopic study of an even lower luminosity
system, Segue\,1.  The systemic velocity of Segue\,1 is far removed
from the Milky Way foreground and thus should be a cleaner object to
study the properties of the least luminous ultra-faint systems.

Segue\,1 was discovered by \citet{belokurov06b} as an over-density of
resolved stars in the SDSS located at ($\alpha_{2000}, \delta_{2000})$
= (10:07:03, +16:04:25) = $(151.763^{\circ}, 16.074^{\circ}$).  Via
isochrone fitting, these authors estimate a distance of
$23\pm2$\,kpc and an absolute luminosity of $M_V \sim -3\pm0.6$.
\citet{martin08a} recently revised the luminosity of Segue\,1 to $M_V
= -1.5^{+0.6}_{-0.8}$ using a more robust method to estimate flux in
systems with small numbers of observable stars.
While the possibility of tidal tails and/or tidal distortion of
Segue\,1 was found in the initial SDSS analysis, deeper imaging and
more thorough simulations suggest that these features can be explained
via Poisson scatter of the few bright stars in this system
\citep{martin08a}.  Segue\,1 has no detected gas content, with an
observed HI gas mass limit of less than $13\Msun$ \citep{putman08a}.
This limit is consistent with other dSphs around the Milky Way in
which any gas has been presumably removed via ram pressure stripping
or used up via tidally-induced star formation \citep{mayer06a}.

\citet{belokurov06b} note that Segue\,1 is spatially superimposed on
the leading arm of the Sagittarius stream.  Because it has a
similar luminosity and size as the most diffuse globular cluster, they
proposed that Segue\,1 is a globular cluster formerly associated with
the Sagittarius dSph.  Spectroscopy of member stars in Segue\,1 is
required to test this hypothesis and answer the crucial question of
whether or not this intrinsically faint stellar system is truly a
globular cluster (i.e.~a stellar system with a single stellar
population with no dark matter).  Here, we present Keck/DEIMOS
multi-object spectroscopy for individual stars in the vicinity of
Segue\,1, identifying \nmembers\ stars as members of Segue\,1.

This paper is organized as follows: in \S\,\ref{sec_data} we discuss
target selection and data reduction for our Keck/DEIMOS spectroscopy.
In \S\,\ref{sec_kin} we discuss the spectroscopic results
including estimates of the velocity dispersion, mass, mass-to-light
ratio and metallicity.  In \S\,\ref{sec_sgr}, we examine the spatial
and kinematic position of Segue\,1 relative to the Sagittarius stream.
In \S\,\ref{sec_gamma} we note that Segue\,1 may be a good target for
indirect detection of dark matter.  Finally, in \S\,\ref{sec_disc}, we
discuss Segue\,1 in context of the Milky Way dSph population.

Throughout the analysis, we use the photometric properties of Segue\,1
as derived by \citet{martin08a} of $M_V = -1.5^{+0.6}_{-0.8}$ (i.e.\
the 1$\sigma$ magnitude limits are $M_V = -0.9$ and $-2.3$)
and $r_{\rm eff}=4.4'^{+1.2}_{-0.6} = 29^{+8}_{-5}$\,pc.  We also assume
a fixed reddening to Segue\,1 based on the \citet{schlegel98a} value
of E(B-V) = 0.032~mag.  We list these and other key parameters in
Table~1.

\begin{figure*}[t!]
\epsscale{1.0}
\plotone{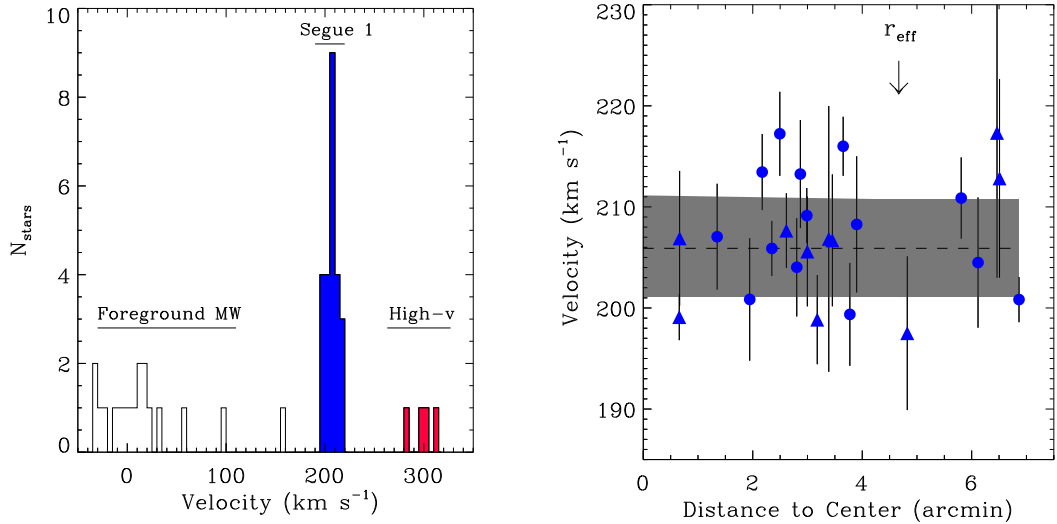}
\caption{{\it Left:\/} Color-magnitude diagram of all stars (small
  black points) within $30'$ of the center of Segue\,1 from SDSS DR~6
  $g$- and $r$-band photometry.  The larger symbols indicate stars
  with measured Keck/DEIMOS velocities: solid blue circles fulfill our
  requirements for membership in Segue\,1, red asterisks are
  higher velocity stars and open squares are foreground Milky Way
  stars.  Two fiducial isochrone are shown shifted to the distance of
  Segue\,1: M92 ([Fe/H] =$-2.3$, solid line) and M3 ([Fe/H]=$-1.6$,
  dashed line).  {\it Right:\/} Spatial distribution of stars near
  Segue\,1.  The solid ellipse is the half-light radius of
  Segue\,1 as measured by \citet{martin08a}.\label{fig_cmd}}
\end{figure*}

\begin{figure*}[t!]
\plotone{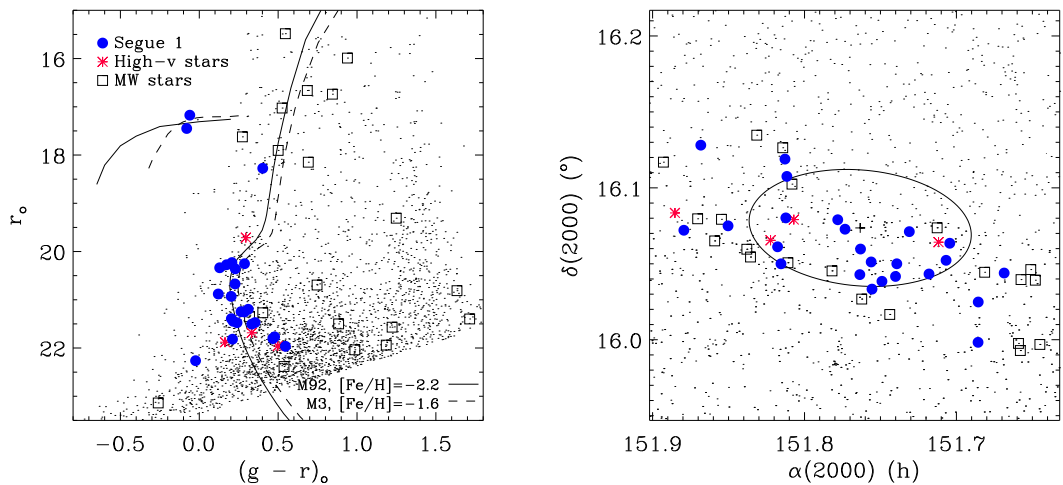}
\caption{{\it Left:\/} Keck/DEIMOS velocity histogram for all stars in
  our sample; velocities are corrected to the heliocentric frame.  We
  identify Segue\,1 as the velocity peak near $v=206$\kms.  Stars with
  less positive velocities are identified as foreground Milky Way, the
  four stars with $v\sim300$\kms\ are tentatively associated with the
  Sagittarius stream as discussed in \S\,\ref{ssec_highv}.  {\it
    Right:\/} Radial distance from the center of Segue\,1 plotted
  against heliocentric velocity.  Stars to the East of the galaxy
  center are plotted as triangles, stars to the West are plotted as
  squares.  We indicate the effective half-light radius ($r_{\rm
    eff}$), the mean systemic velocity of the system (black dashed
  line) and velocity dispersion (grey shaded region). \label{fig_velocity}}
\end{figure*}

\section{Data}\label{sec_data}

\subsection{Target Selection}\label{subsec_targets}

Targets were selected for spectroscopy based on $gri$ photometry of
Segue\,1 from the SDSS DR6 public database \citep{dr6}.  As discussed
in SG07, we set the target priorities to preferentially observe stars
with a high likelihood of being Segue\,1 members.  Using the
theoretical isochrones of \citet{clem08a} and \citet{girardi04a}, we
chose targets whose color and apparent magnitudes minimize the
distance from the best fitting Segue\,1 isochrone.  The highest
priority targets were those located within 0.1~mag of the RGB tracks,
or within 0.2~mag of the horizontal branch, with additional preference
being given to brighter stars (Figure~\ref{fig_cmd}).  Stars farther
from any of the fiducial sequences were classified as lower priority
targets.  We designed the slitmask so as to maximize the number of
high priority targets: a total of 59 targets were placed on the
Segue\,1 mask, 26 of which were in our highest priority category.
Slitmasks were created using the DEIMOS {\tt dsimulator} package in
IRAF.

\subsection{Spectroscopy and Data Reduction}\label{subsec_redux}

One multislit mask was observed for Segue\,1 using the Keck~II 10-m
telescope and the DEIMOS spectrograph \citep{faber03a} on the night of
November 12, 2007.  The mask was observed for a total of 5400~seconds
through the 1200~line~mm$^{-1}$ grating covering a wavelength region
$6400-9100~\mbox{\AA}$.  The spatial scale is $0.12''$~per pixel, the
spectral dispersion of this setup is $0.33~\mbox{\AA}$, and the
resulting spectral resolution is $1.37~\mbox{\AA}$ (FWHM). Slitlets
were $0.7''$ wide.  The minimum slit length was $5''$ to allow
adequate sky subtraction; the minimum spatial separation between slit
ends was $0.4''$ (three pixels).

Spectra were reduced using a modified version of the {\tt spec2d}
software pipeline (version~1.1.4) developed by the DEEP2 team at the
University of California-Berkeley for that survey. A detailed
description of the two-dimensional reductions can be found in SG07.
The final one-dimensional spectra are rebinned into logarithmic
wavelength bins with 15\,\kms\ per pixel.

\subsection{Radial Velocities and Error Estimates}\label{subsec_rvel}

Radial velocities were measured by cross-correlating the observed
science spectra with a set of high signal-to-noise stellar templates.
The method is the same as that described in SG07 and briefly repeated
here.  Stellar templates were observed with Keck/DEIMOS using the same
setup as described in \S\,\ref{subsec_redux} and covering a wide range
of stellar types (F8 to M8 giants, subgiants and dwarf stars) and
metallicities ([Fe/H] = $-2.12$ to $+0.11$~dex).  We calculate and
apply a telluric correction to each science spectrum by cross
correlating a hot stellar template with the night sky absorption lines
following the method in \citet{sohn06a}.  The telluric correction
accounts for the velocity error due to mis-centering the star within
the $0.7''$ slit caused by small mask rotations or astrometric errors.
We apply both a telluric and heliocentric correction to all velocities
presented in this paper.

The random component of the velocity error is calculated using a Monte
Carlo bootstrap method.  Noise is added to each pixel in the
one-dimensional science spectrum, we then recalculate the velocity and
telluric correction for 500 noise realizations.  The random error is
defined as the square root of the variance in the recovered mean
velocity in the Monte Carlo simulations.  The systematic contribution
to the velocity error was determined by SG07 to be 2.2\kms\ based on
repeated independent measurements of individual stars.  Since we are
using the same spectrograph setup and reduction methods, we assume the
systematic error contribution is constant across the two runs.  We add
the random and systematic errors in quadrature to arrive at the final
velocity error for each science measurement.  Radial velocities were
successfully measured for 49 of 59 extracted spectra.  The median
velocity error of these 49 stars is 3.6\kms\, similar to that of SG07.
The median velocity error of the 24 Segue\,1 members (see below) is
5.2\kms\ since these stars are fainter than the sample average.  The
majority of spectra for which we could not measure a redshift did not
have sufficient signal-to-noise.  The fitted velocities were visually
inspected to ensure reliability.  The resulting velocities and
associated errors are listed in Table~2.

\section{Spectroscopic Observations of Segue 1} \label{sec_kin}

\subsection{Foreground Contamination and Membership Criteria} \label{ssec_members}

In Figure~\ref{fig_velocity}, we identify Segue\,1 as the over-density
of stars with radial velocities near 206\kms.  We estimate possible
foreground contribution below and then discuss our criteria for
Segue\,1 membership, which we base only on velocity.

We expect minimal contamination from foreground Milky Way stars at the
position and velocity of Segue\,1. Segue\,1 lies at a Galactocentric
position of $(l,b) = (220.5^{\circ}, 50.4^{\circ})$.  According to the
Besancon starcount model\footnote{http://model.obs-besancon.fr} of the
Milky Way \citep{robin03a} at this Galactic position, the velocity
distribution of Milky Way foreground stars peaks at a heliocentric
velocity of 20\kms.  The Besancon models include stellar contributions
from the Milky Way thin and thick disk, spheroid and stellar halo.
The kinematic distribution of foreground stars is roughly approximated
by a Gaussian with FHWM of 35\kms, however the tails of the
distribution extend to significantly positive and negative velocities.
The percentage of Milky Way stars expected in the presumed velocity
span of Segue\,1, between $190 < v < 220$\kms, is 2.5\% of the total
distribution.  Thus, if we assume that all the stars with velocities
less than 100\kms\ are Milky Way foreground stars (a total of 20
stars, see Figure~\ref{fig_velocity}), we predict less than one
foreground star in the Segue\,1 velocity range.

As noted by \citet{belokurov06b}, Segue\,1 is superposed on the
leading arm of the Sagittarius stream, $\sim100$ degrees away from the
main body of the Sagittarius dSph.  Thus, a second possible source of
contamination in our Segue\,1 sample are stars associated with the
Sagittarius stream.  As discussed in \S\,\ref{sec_sgr}, the predicted
velocity of the leading stream at this position is $v
\sim -100$\,\kms, very far from the radial velocity of Segue\,1.
While both the trailing arm and possible older wraps of the
Sagittarius stream may be present at this position, both components
would have much wider velocity distributions than Segue\,1.  We
conclude that there is no contamination from Sagittarius stream stars
in the Segue\,1 velocity window.  Finally, there are four stars at
higher velocities ($v \sim 300$\kms) that do not appear to be
associated with Segue\,1 as a result of the 100\,\kms\ velocity
difference which we discuss in \S\,\ref{ssec_highv}.

Since the expected contamination from both foreground Milky Way stars
and the Sagittarius stream is low in the velocity region of Segue\,1,
our criteria for Segue\,1 membership is simple: we assign membership
based only on velocity.  Stars with radial velocities between $190 < v
< 220$\,\kms\ are considered members of Segue\,1.  This cut provides
\nmembers\ member stars.  The nearest non-members in our
spectroscopic sample are at $v = 155$\,\kms\ and $v = 281$\,\kms, so
different velocity cuts would identify the same set of members.

The color-magnitude distribution of kinematically-selected Segue\,1
members is shown in Figure~\ref{fig_cmd}.  We plot fiducial sequences
for the globular clusters \object{M92} ([Fe/H] = $-2.3$) and
\object{M3} ([Fe/H]=$-1.6$).  These ridgelines are based on those of
\citet{clem08a} in $g'-r'$, converted to $g-r$ using the
transformations of \citet{rider04a} and shifted to the distance of
Segue\,1 (23\,kpc).  These fiducials are well-matched to the kinematic
sample.  In particular, the spectra of the two bright blue stars
($r\sim17.5$, $g-r \sim-0.1$) show strong broad absorption lines of
the Paschen series and narrow Ca II triplet lines consistent with
being horizontal branch stars.  The position of these two stars is
also well matched to the metal-poor horizontal branch isochrones at
the distance of Segue\,1.

\begin{figure}
\epsscale{0.9}
\plotone{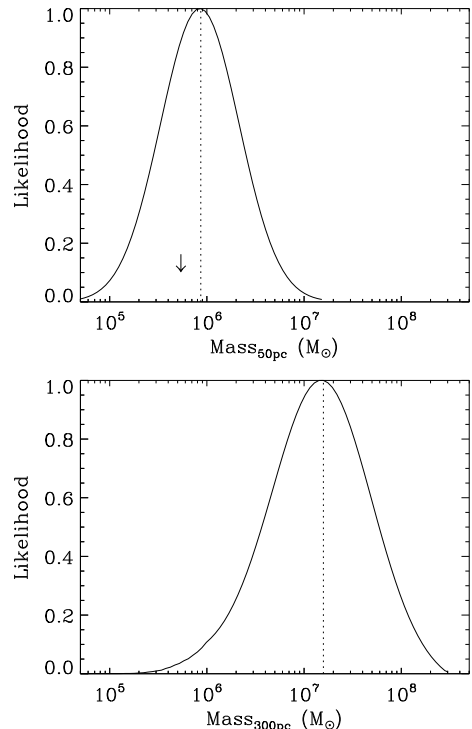}
\caption{Likelihood distributions for the mass of Segue\,1 enclosed
  within 50\,pc ({\it top}) and 300\,pc ({\it bottom}) determined using
  a two-component model as described in \S\,\ref{ssec_mass} and
  \citet{strigari08a}.  The arrow on the top plot indicates the mass
  of Segue\,1 assuming mass-follows-light. Dotted lines show the
  best-fitting two-component model mass and $1-\sigma$
  errors.\label{fig_maxlike}}
\end{figure}

\subsection{Velocity Dispersion}\label{ssec_vdisp}

We measure the mean velocity and velocity dispersion of Segue\,1 using
the maximum-likelihood method described by \citet{walker06a}.  This
method assumes that the observed velocity dispersion is the sum of the
intrinsic galaxy dispersion and the dispersion produced by measurement
errors.  Fitting the full Segue\,1 sample based on the \nmembers\
member stars identified above, we find a mean heliocentric velocity of
$206.4\pm1.3$\kms\ and a velocity dispersion of $4.3\pm1.2$\kms
(Figure~\ref{fig_velocity}).  We do not find evidence for rotation in
this system, however, given the small numbers of stars we cannot rule
out rotation velocities on the same order as the velocity dispersion.
We test this by adding a sinusoidal term to the systemic velocity,
varying the amplitude and scale radius \citep{strigari08a}. The most
likely value for the rotation amplitude is zero, with an upper
1-$\sigma$ limit of 5\kms. While this test justifies the mass modeling
we use with no streaming motion in the velocities, larger kinematic
data sets and smaller velocity uncertainties are necessary to test
more conclusively for streaming motion Segue\,1.  

The grey shaded region in right panel of Figure~\ref{fig_velocity}
indicates the $1\sigma$ width of the Segue\,1 velocity distribution.
We note that all the member stars lie within $2.5\sigma$ of the
systemic Segue\,1 velocity.  The next nearest star in velocity space
is over $10\sigma$ away.  We interpret this cold distribution as
evidence that there are no stars {\it currently} in the process of
being tidally stripped from Segue\,1 \citep{klimentowski07a}.  We note
here and in \S\,\ref{ssec_tides}, however, that the lack of outliers
is not sufficient to prove that tidal processes are not affecting our
results \citep[e.g.,][]{munoz08a}.  This interpretation also does not
mean that stars have not been previously stripped from Segue\,1, and
still allows for the possibility that tidal interactions are currently
on-going in the dark matter component of this object.  We discuss this
further in \S\,\ref{ssec_tides}.

\subsection{Mass and Mass-to-Light Ratio}\label{ssec_mass}

We calculate the dynamical mass of Segue\,1 using two different
methods.  In both cases, we assume that Segue~1 is a relaxed,
self-gravitating, spherically-symmetric system with no rotational
motion.  We presently ignore any effects on the mass estimates due to
tidal interactions between Segue\,1 and the Milky Way, leaving that
discussion to \S\,\ref{ssec_tides}.  We first assume the simplest
possible configuration: an isotropic sphere in which
mass-follows-light.  Further assuming that the density is described as
a King model and is in virial equilibrium, \citet{illingworth76a}
showed that the total mass is then $M = 167 \beta r_{c}
\sigma^{2}$ where $\beta$ is a parameter that depends on the
concentration of the system and is generally assumed to be 8 for dSphs
\citep{mateo98a}, $r_{c}$ is the \citet{king66a} profile core radius,
and $\sigma$ is the observed average velocity dispersion.  We convert
the measured half-light radius of Segue\,1 to King core radius as $r_c
= 0.64* r_{\rm eff} = 18.6^{+5}_{-3}$\,pc.  Using this method, we
estimate the total mass of Segue\,1 to be $4.5^{+4.7}_{-2.5} \times
10^5$\,\Msun.

Our second method to calculate the mass loosens the constraint that
mass-follows-light and uses the individual stellar velocity
measurements (in contrast to the velocity dispersion averaged over the
projected radius as above). The method is described in Strigari et
al.~(2008a).  Similar to the mass-follows-light method, this model
assumes spherical symmetry and dynamical equilibrium, i.e.~that the
kinematic tracer population is related to the mass distribution via
the Jeans Equation.  We assume that the light profile follows the
observed Plummer profile with effective radius $r_{\rm eff} = 29$\,pc,
and that the dark matter follows a five-parameter density profile
characterized by a scale density, a scale radius, an asymptotic inner
slope, an asymptotic outer slope, and a parameter governing the
transition between these two slopes.  The dark matter density profile
allows for both flat central density cores and steep central density
cusps, including the CDM-favored NFW-like $r^{-1}$ central cusps. We
also allow for a radially-varying stellar velocity anisotropy profile.
We then marginalize over these five free parameters, and can estimate
the mass at any given radius.  Though the data do not constrain any of
these parameters separately, the total dynamical mass within the
stellar extent of 50\,pc is relatively well-constrained.  Using this
model, we find the mass within 50\,pc to be $8.7_{-5.2}^{+13} \times
10^5$M$_\odot$.  We note that the likelihood distribution of this
quantity, shown in Figure~\ref{fig_maxlike}, is nearly log-normal; the mass is
greater than $5\times10^4$\Msun with 3-$\sigma$ confidence.  In
comparison, the total stellar mass is merely 340\Lsun.


The two masses calculated above agree within errors. In the
mass-follows-light method, the majority of dark matter mass in the
galaxy resides within the stellar radius, while the second
method leaves open the possibility that the majority of the mass lies
{\em outside} the observed light distribution. However, determining
the total galaxy mass requires knowledge of the total radial extent of
the dark matter halo and the profile shape beyond the last observed
point.  This is clearly difficult to determine observationally and
strongly depends on the unknown orbital history of the galaxy.  We can
only estimate the instantaneous tidal radius of the galaxy, which
ranges from a few ten to a few hundred parsecs depending on
assumptions detailed in \S\,\ref{sssec_tides}.  If the tidal radius is
large, then it is plausible that the majority of mass lies outside the
stellar distribution.  

Extrapolating the second estimate of mass to a
radius of 300\,pc, we find a total dynamical mass of $10^7$ M$_\odot$,
which, remarkably, is consistent with the common mass scale of all
Milky Way dSphs as seen in Figure~\ref{fig_corr} \citep{strigari08a}.
This common mass scale has been noted in previous studies
\citep{mateo93a, gilmore07a}; we discuss this further in
\S\,\ref{sec_disc}.  If the stellar component of Segue\,1 is embedded
in a $10^7$\Msun\ dark matter halo, we would expect the luminous
component to have experienced very little tidal disruption despite its
current proximity to the Galaxy.

Regardless of which estimator is used above, the observed mass of
Segue\,1 is {\it significantly} larger than expected if all its mass
were due to a stellar-only component.  Since Segue\,1 contains little
to no HI gas \citep{putman08a}, the stellar mass likely dominates the
total baryonic mass.  In the absence of non-baryonic dark matter, we
expect the mass-to-light ratio of Segue\,1 to be $M/L_V\sim 3$,
accounting for stellar remnants in an old stellar population
\citep{maraston05a}.  Assuming this mass-to-light ratio, the stellar
mass of Segue\,1 is $\sim 1\times 10^3$\Msun, translating into an
expected velocity dispersion of merely $0.4$\kms.  This is more than
3-$\sigma$ below the measured dispersion of Segue\,1 and thus argues
strongly for the presence of dark matter.

Finally, we calculate the $V$-band mass-to-light ($M/L_V$) ratio
within the observed radius.  Combining the absolute luminosity of
Segue\,1 ($M_V = -1.5^{+0.6}_{-0.8}$) with the mass from the first
method above (assuming mass-follows-light), we calculate a
mass-to-light ratio of ln$(M/L_V) = 7.2^{+1.1}_{-1.2}$ ($M/L_V =
1320^{+2684}_{-936}$), and in the second two-component method we
calculate ln$(M/L_V) = 7.8^{+0.5}_{-1.3}$ ($M/L_V =
2440^{+1580}_{-1775}$).  In both cases, the error distribution is
asymmetric and the mass-to-light ratio is well in excess of that
predicted from the stellar mass alone.  The two-component model ratios
suggests a dark matter-dominated galaxy with a 6$-\sigma$
significance.  If the luminous components of dSphs do indeed reside in
common mass dark matter halos, we would predict the highest M/L ratios
in the least luminous dSphs (see middle panel Figure~\ref{fig_corr}).
Since Segue\,1 is the least luminous of the recently discovered
ultra-faint Milky Way satellites, this remarkably high M/L is expected
in this model.  Understanding the processes that lead to this high M/L
will be a future challenge to galaxy formation models.

\subsection{Possible Caveats on the Mass of Segue 1}\label{ssec_tides}

The remarkably high mass-to-light ratio of Segue\,1 rests on our
interpretation that the measured stellar velocities faithfully trace
the gravitational potential.  Here we discuss two possibilities that
might affect this assumption.  First is the presence of unresolved
binary stars inflating our measured velocity dispersion.  The second
is tidal interactions with the Milky Way affecting the kinematics.
Both issues are difficult to quantify without further observations,
but are worthwhile considering here.

\subsubsection{The Effects of Binary Stars}\label{sssec_binaries}

If a high percentage of our Segue\,1 stellar members are in fact
unresolved binary star systems, the measured velocity dispersion may
be inflated due to their orbital motion.  The severity of this effect
depends on the mass ratio of individual systems, with equal mass
binaries contributing most to the velocity bias.  The likely presence
of binaries in our kinematics sample is difficult to estimate
empirically without repeated velocity measurements.
\citet{olszewski96a} simulate the effects of binaries on the velocity
dispersions of dSph with datasets somewhat similar to that of
Segue\,1.  Assuming a solar neighborhood binary fraction, they suggest
that the velocity dispersion due to binaries alone is on the order
$\sim1.5$\,\kms.  Thus, while the overall dispersion may be inflated
by binaries, they cannot explain away the strong evidence of dark
matter.  Because the true binary fraction in Segue\,1 may be very
different than that of the solar neighborhood, we do not fold this
systematic error into our mass estimates.  Additional observations and
improved simulations will allow us to better quantify the effects of
binaries on our results.

\subsubsection{The Effects of Tidal Interactions}\label{sssec_tides}

It is likely that Segue\,1 has been affected by tidal interactions
with the Galaxy.  Segue\,1 lies at a distance of 23\,kpc from the Sun,
or d$_{\rm GC-Seg1}=28$\,kpc from the Galactic Center (GC) assuming a
GC-Sun distance of 8.5\,kpc.  We estimate the instantaneous tidal
radius for Segue\,1 by first approximating both the Milky Way and
Segue\,1 as point masses.  We assume a enclosed Milky Way mass of
$3\times 10^{11}\Msun$, corresponding to a circular velocity of
210\kms\ at 28\,kpc.  This is consistent with the Milky Way model used
in \S\,\ref{sec_sgr} and \citet{law05a}.  Using the first estimate for
the mass of Segue\,1 in \S\,\ref{ssec_mass}, we determine a tidal
radius of $r_{\rm t}= (M_{\rm Seg1}/M_{\rm MW})^{(1/2)} d_{\rm
  GC-Seg1} = 33$\,pc.  Assuming that Segue\,1 is embedded in an
extended dark matter halo (using mass from the second method above at
300\,pc), the tidal radius increases to $r_{\rm t}=160$\,pc.  In the
first case, the luminous matter extends beyond the tidal radius and we
would expect to see evidence for unbound stars in our kinematic
sample, in the second case we would expect our observations to be well
within the bound radius.  However, this calculation does not account
for the fact that Segue\,1 is orbiting the Milky Way.  When
calculating the tidal radius in the reference frame of the MW-Segue\,1
system, stars in Segue\,1 will also feel centrifugal and coriolis
forces in this rotating reference frame.  If we include these forces,
the tidal radius (also called the Jacobi or Roche radius, Binney \&
Tremaine (2008), Eqn.\,8.91) is then $r_{\rm t}= (M_{\rm
  Seg1}/3*M_{\rm MW})^{1/3}d_{\rm GC-Seg1}$.  The tidal radius then
increase to 220 and 615\,pc for the first and second cases,
respectively.  These are of course estimates of the {\it
  instantaneous} tidal radius: if Segue\,1 is on an elliptical orbit
the tidal radius may have been much different in the past.

If some, or all, of the stars associated with Segue\,1 are tidally
disrupting (unbound), then the measured velocities likely provide an
inflated estimate of the mass \citep{klimentowski07a}.  In the extreme
case that Segue\,1 is completely unbound, its mass could be as low as
the stellar component alone ($10^3$\Msun).  However, the fact that the
luminosity profile of Segue\,1 is centrally concentrated suggests that
this object is not completely unbound.  The crossing time of Segue\,1
(assuming $r_{\rm eff} = 30$\,pc and a velocity dispersion of 4.3\kms)
is $10^7$\,years.  The travel time along the orbit of Segue\,1 in
$10^7$ years is roughly 2\,kpc assuming a circular orbital speed of
200 \kms.  Thus, we would naively expected an unbound version of
Segue\,1 reside only a few kpc away from its disruption site before
quickly dissipating, making this extreme scenario unlikely.  Detailed
dynamical modeling appropriate to this system and a knowledge of
Segue\,1's orbital history is required to properly determine the
degree of tidal interactions and disruption in this system.  Our mass
estimates presented in this paper explicitly assume that the
kinematics of Segue\,1 are not affected by tides.

To determine the true orbit of Segue\,1, we need to know its
transverse motion.  The heliocentric velocity of Segue\,1 is
$v=206$\kms\ and the velocity relative to the Galactic Standard of
Rest\footnote{To compute velocities in the Galactic standard of rest
  (GSR) frame, we assume the solar peculiar velocity is $(U,V,W) =
  (9,12,7)$ km s$ ^{-1}$ relative to the local standard of rest, for
  which we adopt a rotation velocity of 220 km s$^{-1}$.}  (GSR) is
$v_{\rm GSR}=114$\kms.  We can rule out a circular orbit: in the Milky
Way model discussed in \S\,\ref{sec_sgr}, the maximum projected GSR
velocity for a circular orbit is $v_{\rm GSR, circ} = 55$\kms at the
distance of Segue\,1.  If the transverse motion of Segue\,1 is similar
to or less than that of the measured radial motion, then Segue\,1 was
closer to the Galactic Plane in the recent past.  However, there is no
clear evidence to suggest tidal stripping is currently affecting the
luminous component.  We do not see velocity outliers in our kinematic
sample (which might indicate that these stars are in the process of
being stripped), nor other clear evidence of on-going tidal disruption
(e.g.~photometric evidence of tidal tails or tidally-induced
rotation).  While the absence of these features cannot be used as
proof that tidal stripping is not on-going \citep{munoz08a}, it is
consistent with our assumption that tidal stripping is not currently
affecting the luminous component.  An estimate of the proper motion of
Segue\,1 is needed to constrain the orbital history of this object.

\begin{figure*}[t!]
\plotone{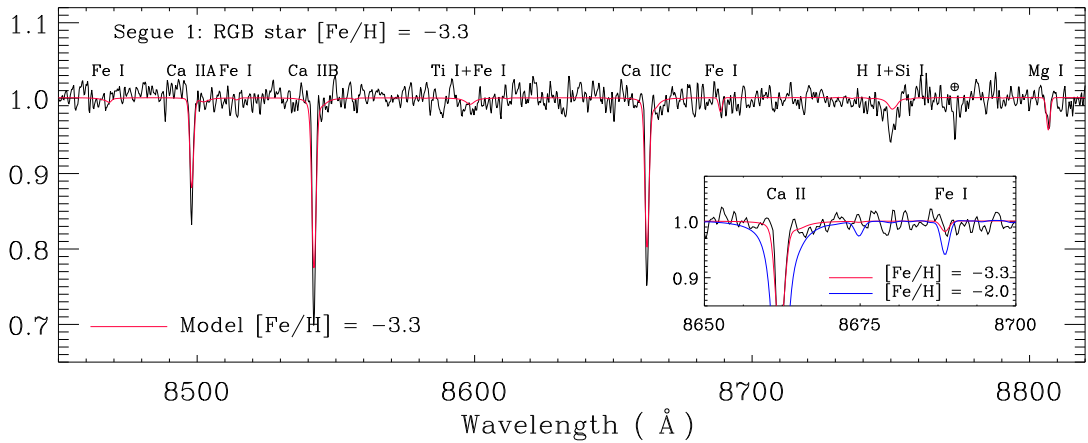}
\caption{Keck/DEIMOS spectrum for the Red Giant Branch star member of
  Segue\,1.  This star has a measured metallicity of [Fe/H] = $-3.3
  \pm 0.2$\,dex as determined via spectral synthesis \citep{kirby08a}.  The model
  spectrum is shown in red. The cores of the \ion{Ca}{2} triplet lines
  are not well-modeled because they form out of local thermodynamic
  equilibrium and do not play any role in the
  metallicity determination. Inset is a zoom-in of a region near the
  Ca II C-line.  The red [Fe/H] model is compared to a more metal-rich model
  ([Fe/H] = $-2.0$, blue line) with the same atmospheric parameters.  The
  observed spectrum shows no evidence for absorption at
  8689~$\mbox{\AA}$, even though a star as metal-poor as [Fe/H] =
  $-2.0$ would display this line.\label{fig_spectra}}
\end{figure*}

\subsection{Metallicity}\label{ssec_metal}

We estimate the spectroscopic metallicity of individual stars in our
Segue\,1 sample via spectral synthesis modeling \citep{kirby08a}.  The
method compares the observed spectrum to a grid of synthetic spectra
covering a range of effective temperature, surface gravity and
composition.  We estimate effective temperature and surface gravity
for each star based on the Johnson-Cousins $VI$ magnitude which we
determine by transforming the SDSS $gri$ magnitudes \citep{chonis08a}.
The results are unaffected by using alternative photometric methods to
determine these parameters.  The best matching composition is found by
minimizing residuals between the observed spectrum and a smoothed
synthetic spectrum matched to the DEIMOS spectral resolution.  Our
method has been tested against high resolution Keck/HIRES abundances
for six RGB stars in the ultra-faint dSphs of SG07 \citep{kirby08b}.
This comparison yields precisions better than 0.25\,dex for DEIMOS
spectra with signal-to-noise ratios (S/N) greater than $\mathrm{S/N} >
20~\mbox{\AA}^{-1}$.  Although the method can theoretically be applied
to all types of stars, it has not yet been tested against
high-resolution spectroscopic abundances for horizontal branch or main
sequence stars.

Our kinematic sample contains a single RGB star ($r=18.4$).  The above
method estimates its metallicity to be [Fe/H] = $-3.3\pm0.2$\,dex.
The effective temperature and surface gravity used to determine the
metallicity of this star are $T_{\rm eff}=5191$\,K and $\log g =
2.76$, with estimated systematic errors of 150\,K and 0.12\,dex,
respectively.  The derived metallicity is much more sensitive to
$T_{\rm eff}$ than $\log g$; \citet{kirby08a} estimate that a 150\,K
change in $T_{\rm eff}$ incurs an error on [Fe/H] of less than
0.15\,dex.  While DEIMOS spectra contain some information about
$\alpha$-element abundances, the errors we estimate on this quantity
are large.  The [Fe/H] value is unchanged whether or not we mask out
absorption lines due to the $\alpha$-elements.

A small portion of the observed Segue\,1 RGB star spectrum and synthetic
spectra are shown in Figure~\ref{fig_spectra}.  At this metallicity,
the strong absorption lines of \ion{Ca}{2} are clearly visible, but
weaker Fe lines are not.  In the inset to Figure~\ref{fig_spectra}, we
compare a small region of the observed spectrum to models at [Fe/H] =
$-3.3$ and $-2.0$ with the same temperature and surface gravity.  For
a star with these parameters, the strongest Fe line in the DEIMOS
spectral range is \ion{Fe}{1}~8689.  The observed spectrum in
Figure~\ref{fig_spectra} shows no evidence for absorption at
8689~$\mbox{\AA}$, even though a more metal-rich star would display
this line.

At red wavelengths, metallicity is often estimated via the \ion{Ca}{2}
triplet absorption lines \citep[e.g.][SG07]{helmi06a}.  However,
\citet{kirby08b}, \citet{koch08a} and others note that current
implementations of this method fail for metallicities below
[Fe/H]$\sim -2.5$.  The \ion{Ca}{2} triplet method is based an
empirical calibration of Galactic globular clusters and is not
calibrated for metallicities below [Fe/H] $\le -2.4$
\citep{rutledge97a}.  The metallicities in the ultra-faint dSph are
below this limit.  We therefore do not use this method and strongly
caution the use of this relationship for very low metallicity systems.
The remaining analysis is based on the results from the spectral
synthesis method above.

While other stars in our kinematic sample have sufficient
signal-to-noise to measure metallicity, [Fe/H] estimates for the
remainder of the sample are less reliable.  The two horizontal branch
stars seen in Figure~\ref{fig_cmd} are too hot to display strong metal
absorption and what metal lines exist are overwhelmed by the Paschen
series.  The main sequence stellar spectra are more suitable for
metallicity measurement, but have much lower signal-to-noise as
compared to the RGB star above and higher surface gravities.  The
synthesis method has also not yet been tested for stars with $\log g >
3.3$.  The main sequence stars with adequate S/N to measure a
metallicity have surface gravities $3.5 < \log g < 4.3$.  The average
metallicity for these thirteen main sequence stars is [Fe/H] =
$-1.8\pm 0.1$\,dex, with individual measurements ranging from $-1.5$
to $-2.8$.  This average is significantly more metal-rich than above
and suggests that the mean metallicity of Segue\,1 may be higher than
that of the single RGB star.  These results also suggest that Segue\,1
has a significant internal metallicity spread.  In support of this
spread, we note that the fiducial isochrones in the color-magnitude
diagram of Figure~\ref{fig_cmd} cannot simultaneously fit the RGB and
main sequence.  While the horizontal branch and main sequence turn-off
are well fit in this figure, the single RGB star is slightly too blue,
suggesting it is more metal-poor than the main sequence, consistent with
our spectroscopically measured metallicity.  These results, however,
should be approached with caution.  While there are no obvious reasons
the main sequence metallicities should be biased, we remain
aware that the spectral synthesis code has not been tested in this
regime.  Pending more reliable confirmation, we take the metallicity
of the RGB star to be representative of Segue\,1.

Kirby et al.\ (2008b) demonstrate that the luminosity-metallicity
relationship is log-linear for Milky Way dwarf galaxies across nearly
four decades in luminosities (see Figure~\ref{fig_corr} and
\S\,\ref{sec_disc}).  Given the luminosity of Segue\,1 ($M_V=-1.5, L_V
= 340\Lsun$), the predicted metallicity based on this relationship is
[Fe/H] = $-2.8\pm0.2$.  While our metallicity estimate of the single
RGB star in our Segue\,1 sample is more metal-poor than this
prediction, the main sequence metallicity is more metal-rich.  The
average of these two metallicities is closer to the predicted value.
Additional observations are required to securely determine whether or
not Segue\,1 lies on the luminosity-metallicity relationship, and in
Figure~\ref{fig_corr} we assume that the average metallicity is equal
to that of the RGB star.  Quantifying the mean metallicity of Segue\,1
and the amount of internal metallicity spread is crucial to
interpreting the formation history of Segue\,1.  If this object does
indeed lie on the luminosity-metallicity relationship and has a
significant internal metallicity spread, this is further evidence for
that Segue\,1 formed via galaxy, rather than globular cluster,
formation processes.

\section{Segue\,1, Distinct from the Sagittarius Stream}\label{sec_sgr}


Segue\,1 is spatially super-imposed on the leading arm of the
Sagittarius (Sgr) stream.  This placement and its tiny size led
\citet{belokurov06b} to identify it as a possible globular cluster
formerly associated with the Sgr dSph.  Six other globular clusters
have been associated with the Sgr stream \citep{bellazzini03a,
  casetti-dinescu07a, carraro07a}.  While our measured
velocity dispersion and inferred M/L ratio of Segue\,1 suggest that it
is not a globular cluster, it is still possible that it could have
been a dwarf satellite of Sgr that has been captured by the Milky Way.
We now investigate whether or not Segue\,1 could be kinematically
associated with the Sgr Stream.

\begin{figure*}
\plottwo{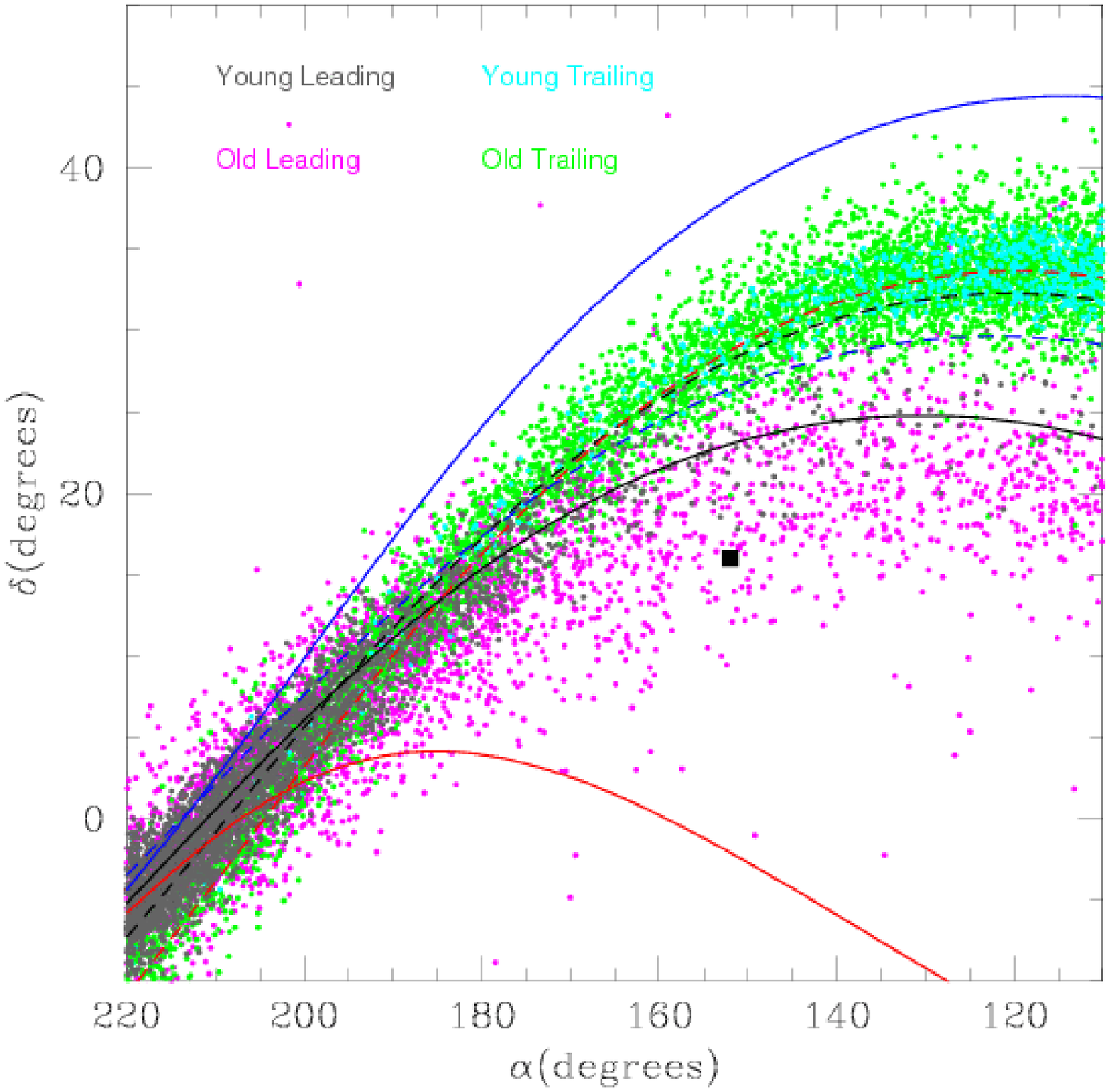}{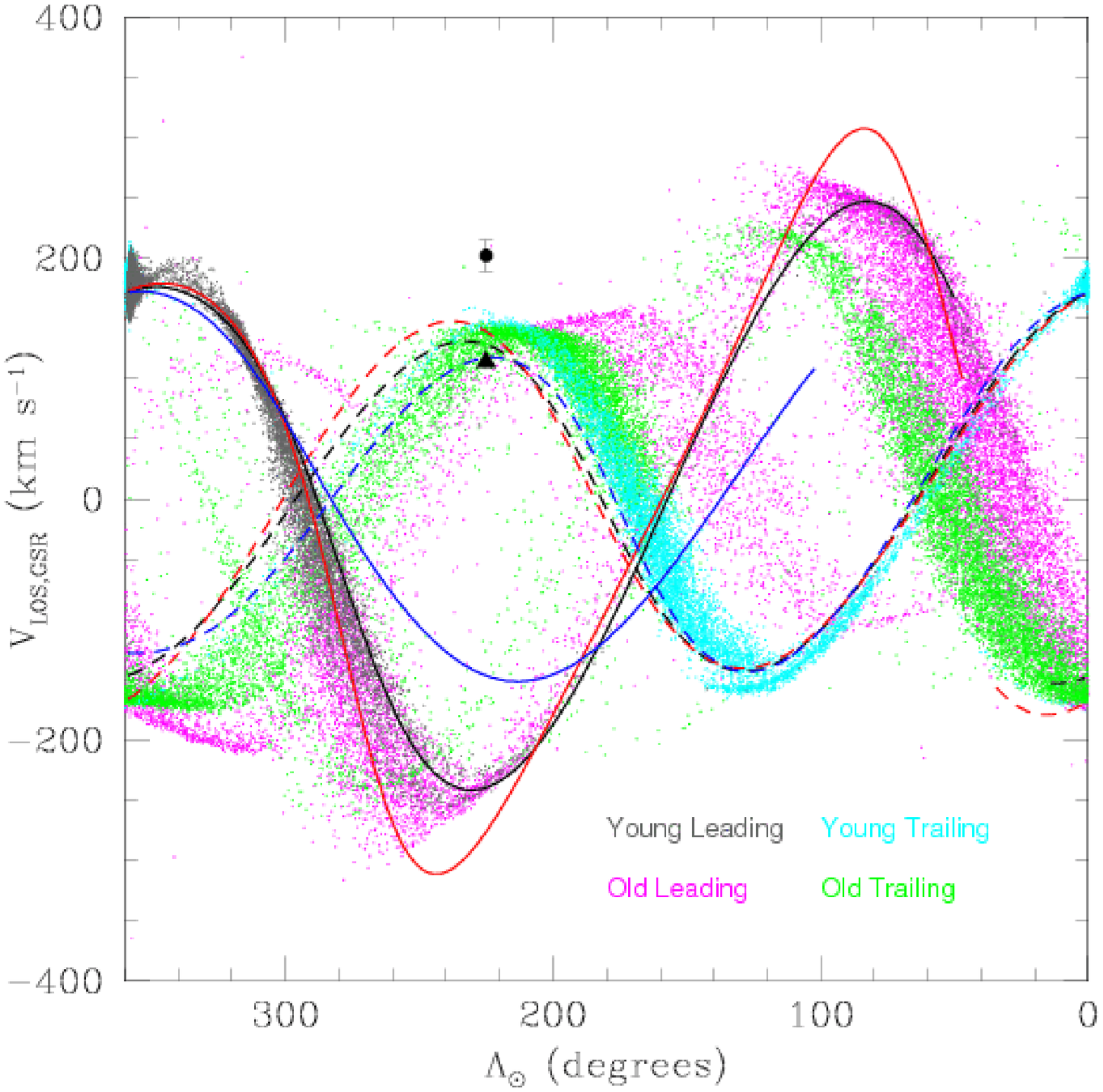}
\caption{N-body model debris (colored points) from the Sgr dSph is
  plotted as a function of ({\it left}) sky coordinates and ({\it
    right}) line-of-sight velocity (relative to the GSR) as a function
  of orbital longitude $\Lambda_{\odot}$ \citep{majewski03a}.
  Grey/cyan points represent debris lost from Sgr during the last 5
  Gyr and magenta/green points are debris lost more than 5 Gyr ago in
  the leading/trailing tidal tails respectively.  Overplotted is the
  orbit of the Sgr dwarf core (black line) shown as a solid/dashed
  line for leading/trailing portions of the orbit respectively.
  Red/blue lines respectively represent the orbits of the $q = 0.9$
   and 1.25 and models of \citet{law05a} for comparison.  The
  location of the Segue\,1 field is indicated by a square in the left
  panel, and the velocity of the Segue I dwarf (triangle) and
  high-velocity feature (circle) are indicated in the right hand
  panel.  The error bars on the high-velocity feature indicate the
  1$\sigma$ spread about the mean value of stars in the feature.}
\label{sgr.fig}
\end{figure*}

\citet{majewski03a} defined a longitudinal coordinate system,
$\Lambda_{\odot}$, in which the center of the Sgr dSph lies at
$\Lambda_{\odot}=0$.  In this system, Segue\,1 is roughly
130$^{\circ}$ away from the main body of the Sgr dSph at
$\Lambda_{\odot} = 224.5^{\circ}$ (Figure~\ref{sgr.fig}).  Unlike the
region near the Sgr dSph or the trailing stream of recent tidal debris
($\Lambda_{\odot} \sim 0 - 100^{\circ}$), the kinematics of the stream
in the region near Segue\,1 are not well determined observationally.
We therefore compare our data to numerical N-body models in order
to determine whether Segue\, may be kinematically as well as spatially
associated with Sgr tidal debris.

Our model of the Sgr stream is similar in many respects to the $q =
1.0$ model (i.e.,~that in which the Galactic dark halo potential is
spherical) described by \citet{law05a}, with some modifications made
in order to simultaneously match both the trailing arm M-giant
velocities \citep{majewski04a} and the newly observed SDSS leading arm
bifurcation \citep{belokurov06c}.  In brief, the Galactic halo
flattening in this model is mildly prolate ($q = 1.05$), and the model
Sgr dwarf has an initial mass of $10^{8} M_{\odot}$, a scale length of
350\,pc, and has been orbiting in a static Galactic potential for
$\sim 9$ Gyr.  We refer the reader to \citet{law05a} for a more
thorough discussion of the N-body modeling technique.

Figure~\ref{sgr.fig} (left panel) illustrates the previously noted
spatial alignment of the Segue\,1 field with the leading Sgr stream
(i.e. the `A' and `C' streams of \citet{Fellhauer06a}).  As
demonstrated in the right hand panel, however, the velocity relative to
the GSR of Segue\,1 ($v_{\rm Seg1, GSR} = 114$ km s$^{-1}$) is wildly
discrepant with the leading tidal stream at the corresponding angular
position ($\sim -250$ km s$^{-1}$, grey/magenta points).  Instead,
Segue\,1 appears to be more consistent in velocity with the {\it
  trailing} stream, from which it is offset by $\sim$ 15$^{\circ}$
($\sim 6$\,kpc at the distance of Segue\,1).  {\it Given these
  conflicting data, we conclude that Segue\,1 is not physically
  associated with either stream.}

We note for completeness however that Segue\,1 is consistent in both
angular coordinates and radial velocity with an extremely old segment
of leading arm tidal debris (released from Sgr $\sim 7-8$ Gyr ago)
which has been wrapped roughly 520$^{\circ}$ around the Milky Way from
the Sgr core (i.e. the segment of magenta debris at $\Lambda_{\odot}
\sim 220^{\circ}$ and $V_{\rm LOS,GSR} \sim 100$ km s$^{-1}$).
However, conclusive observational evidence for the existence of such
old, multiply-wrapped tidal debris from Sgr has not yet been
established.  Our models therefore leave open the possibility that
Segue\,1 was initially associated with the Sgr dSph, but was removed
very early in the tidal interactions between Sgr and the Milky Way.
Previous claims of associated cluster systems
\citep[e.g.~][]{bellazzini03a,casetti-dinescu07a} have focused only on
relatively recent debris.

There is of course still considerable uncertainty in the `best' model
for Sgr.  Only models with strongly prolate ($q=1.25$) halos produce
streams that match the leading arm M-giant velocities \citep{law05a},
while models with oblate halos ($q = 0.9$) best match the observed
precession of the M-giant orbital plane \citep{johnston05a}.  In
contrast, a near spherical model is required in order to match the
bifurcated stream observed in the SDSS \citep{belokurov06c}, as is an
extremely low satellite mass ($10^8 M_{\odot}$) which in turn produces
streams too dynamically cold to reproduce the observed dispersion
among the M-giant velocities \citep{law05a}.  In Figure~\ref{sgr.fig} we
demonstrate the behavior of Sgr debris in oblate, near-spherical, and
prolate Galactic dark halo potentials via point-particle orbital
tracks (red, black, and blue lines respectively).  While debris from
an N-body satellite is not perfectly traced by the orbit of the
satellite core, this orbital track gives a good sense of the behavior
of the debris (note how the black line roughly tracks the colored
points) and indicates that it is not possible to construct a model in
which leading tidal debris can match the observed velocity of
Segue\,1.  Similarly, for no model are trailing Sgr debris spatially
coincident with Segue\,1 while simultaneously reproducing the trend of
trailing M-giant velocities observed by \citet{majewski04a}.  While
puzzles obviously remain, these uncertainties do not affect our
conclusion that Segue\,1 cannot be associated with {\it recent} Sgr
debris.

\subsection{Higher Velocity Stars:  An Old Piece of the Sagittarius Stream?}\label{ssec_highv}

There are four stars in our kinematic sample with $v\sim300$\kms, or
$v_{\rm GSR} \sim 200$\kms (Figure~\ref{fig_velocity}).  This is
unusual in that standard Milky Way models predict that such high velocity
stars are extremely rare \citep[e.g.][]{robin03a}.  These four stars
have sufficiently different velocities ($\Delta v = 30$\kms) that they
are not gravitational bound to each other; however, given their
spatial and kinematic proximity they could plausibly be associated
with a single stellar stream.  To highlight how unusual this grouping
is, we note that over the eight fields observed by SG07 (with similar
targeting priorities), only 7 out of nearly 900 stars had $v_{\rm GSR}
\ge 200$\kms, with only one field having more than one higher velocity
stars (as compared to 4 of 59 in Segue\,1).  Since none of the SG07
fields are near any known streams, we circumstantially associated
these four stars with the Sgr stream.  However, none of the Sgr models
discussed above match the position and velocity of these higher
velocity stars (filled circle in Figure~\ref{sgr.fig}).  We
tentatively associate these stars with older Sgr tidal debris or a
possibly a new stream.  More observations and theoretical work is
needed in this region to confirm this hypothesis.

\begin{figure*}
\epsscale{0.8}
\plotone{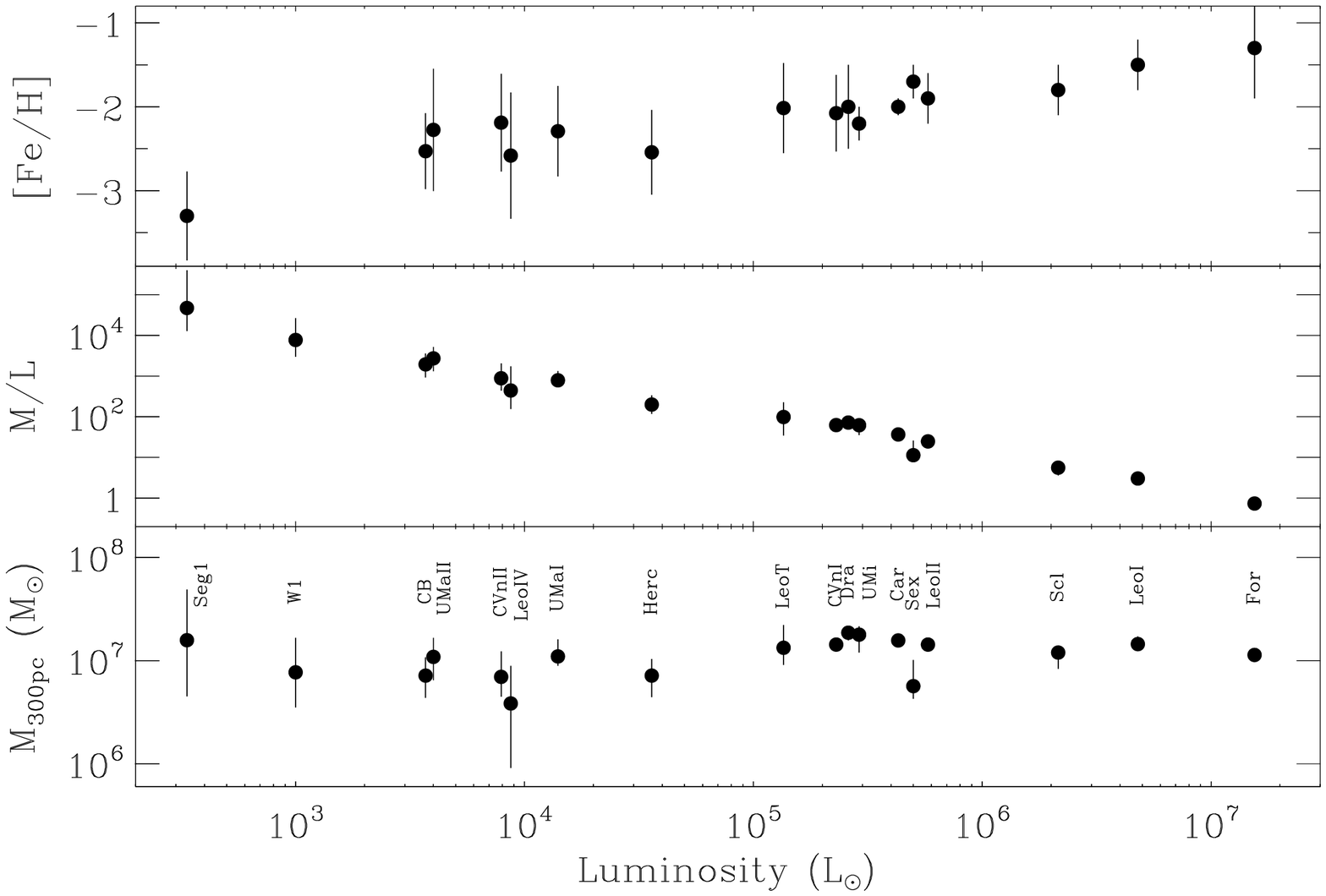}
\vskip 0.2cm
\caption{\small{Segue\,1 lies at the low luminosity end of the
    metallicity([Fe/H])-luminosity, mass-to-light-luminosity and
    mass-luminosity relationships established by the Milky Way
    dSphs. Masses for the Milky Way dSph are taken from
    \citet{strigari08a}, luminosities from \citet{martin08a} and
    \citet{mateo98a}, and metallicities from \citet{kirby08b}.  While
    the luminosity spans nearly five orders of magnitude, the enclosed
    (300\,pc) remains nearly constant at $10^7$\Msun.  Over the same
    luminosity range, the metallicity decreases nearly 2\,dex.  
    Explaining the mechanisms that set these relationships is key to
    understanding galaxy formation at the smallest
    scales.}}\label{fig_corr}
\end{figure*}

\section{The Predicted Gamma-ray Flux from Segue\,1}\label{sec_gamma}

Having established that Segue\,1 is dark matter-dominated, it is
interesting to consider the implications of having a massive dark
matter halo in such close proximity to the Sun.  Generically, dSphs
have been attractive targets for indirect dark matter detection
experiments, via particle annihilation production of gamma-rays, due
to their high dark matter densities and lack of internal
gamma-ray sources~\citep{baltz00a, tyler02a}.  \citet{strigari07b}
note that the even higher dark matter densities of the recently
discovered ultra-faint dSphs, combined with their proximity, make them
particularly interesting candidates for indirect detection. 
Upper limits on the gamma-ray fluxes have so far been reported
for several classical dwarfs including the Draco, Ursa Minor, and
Sagittarius dSph \citep{aharonian08a,driscoll07a,wood08a}.  The recent
launch of Gamma-ray Large Area Telescope (GLAST) satellite
\citep{ritz07a} makes this a particularly timely calculation.

Based on the mass estimates of \S\,\ref{ssec_mass}, the average dark
matter density of Segue\,1 is 1.65\Msun/pc$^{-3}$ inside 50\,pc.  We
determine the gamma-ray flux from dark matter annihilation by
marginalizing over the unknown halo parameters using a maximum
likelihood analysis similar to that described in \S\,\ref{ssec_mass}.
We assume the most optimistic supersymmetric model for the dark matter
particle, and refer to \citet{strigari07b} for additional details of
the input assumptions.  Because we are considering an annihilation
signal, the gamma-ray flux scales as the square of the central density
and as the inverse square of the distance.  Marginalizing over the
appropriate parameters, we find the predicted gamma-ray flux for
Segue~1 is $\Phi_{\gamma} = 5.5^{+10}_{-3.5} \times 10^{-10}$\,photons
cm$^{-2}$ s$^{-1}$.  The mean value of this flux is more than a factor
of ten larger than that from the classical dSphs and is higher than
that predicted for any of the previous ultra-faint dwarfs
\citep{strigari07b}.  Thus, Segue~1 is an excellent target for the
indirect detection of dark matter via particle annihilation.

\section{Discussion}\label{sec_disc}

As seen in Figure~\ref{fig_corr}, Segue\,1 lies on an extension of the
luminosity-metallicity and luminosity-mass relationships established
by brighter Milky Way dSphs.  While the dSphs span nearly five orders
of magnitude in luminosity, their mass enclosed within 300\,pc remains
nearly constant at $10^7$\Msun \citep{strigari08a}. This common mass
scale has been noted in previous studies \citep{mateo93a, gilmore07a},
but remains a very surprising result given the much larger luminosity
range spanned by the present data.  It strongly suggests the existence
of a characteristic scale in either galaxy formation processes or dark
matter physics.  At the same time, the average metallicities of the
dSphs are correlated with luminosity such that stars in the least
luminous dSph are the most metal-poor \citep{kirby08b}. Segue\,1 is at
the extreme end of these relationships: its luminosity is merely $L =
340$\Lsun, yet its total mass enclosed within 300\,pc is $10^7$\Msun
(projecting the mass model discussed in \S\,\ref{ssec_mass}),
resulting in the highest $M/L_V$ ratio of any known stellar system.
The metallicity for the single RGB star in Segue\,1 is [Fe/H] =
$-3.3$\,dex, one of the most metal-poor stars known in a dSph galaxy.
This metallicity is slightly less than that predicted by the Kirby et
al.~log-linear relationship, however, we note that the average
galactic metallicity may be higher than this single star.

The correlations in Figure~\ref{fig_corr} are the key to understanding
how dSphs form.  While several formation avenues exist to modify the
mass-to-light ratio of dSphs, the added constraint of the
luminosity-metallicity correlation reduces the number of allowable
models.  This correlation rules out a tidal stripping scenario in
which lower luminosity systems initially form as more luminous
galaxies outside the environment of the Milky Way and are then tidally
stripped to their present state as they enter the Milky
Way environs.  In this scenario, the metallicity of stars would not be
tied to the present luminosity. While ruling out formation scenarios
is certainly progress, determining what formation processes can
explain the observed correlations will be more challenging
\citep[e.g.][]{bovill08a}.  A key question raised by the Segue\,1
results is why the Milky Way dwarf dSphs have such remarkably
different luminosities, yet appear to have similar total
masses.  Why do all these objects have a common mass halo and is this
consistent with the mass spectrum of dark matter halos predicted by
simulations?  Explaining the mechanism that sets both the
mass-luminosity and luminosity-metallicity relationships in the Milky
Way will provide insight to the formation of galaxies at all scales.

\section{Summary}\label{sec_summ}

Segue\,1 ($M_V = -1.5^{+0.6}_{-0.8}$) is the least luminous of the
ultra-faint galaxies recently discovered in the SDSS, and thus the
least luminous known galaxy.  We present Keck/DEIMOS spectroscopy of
\nmembers\ member stars suggests that Segue\,1 is dark
matter-dominated and metal-poor.  We measure an internal velocity
dispersion of $4.3\pm 1.2$\kms, and infer a total mass of
$4.5^{+4.7}_{-2.5} \times 10^5 M_{\odot}$ in the case where
mass-follow-light; using a two-component maximum likelihood model, we
determine a mass within 50\,pc of $8.7_{-5.3}^{+13} \times
10^5$M$_\odot$.  The two masses agree within errors, however, in the
first case, the majority of dark matter mass in the galaxy resides
within the stellar radius, while the second method leaves open the
possibility that the majority of the mass lies {\em outside} the
observed light distribution.  The metallicity of the single RGB star
in our sample if [Fe/H] = $-3.3$\,dex, well below that of any known
globular cluster \citep{harris96a}.  Although Segue\,1 is spatially
superimposed on the Sagittarius stream, its mean velocity is
inconsistent with recent Sagittarius tidal debris in this region.  Our
models leave open the possibility that Segue\,1 is a dwarf galaxy that
was initially associated with the Sgr dSph, but was stripped away
early in tidal interaction between Sagittarius and the Milky Way.
Finally, we note that the combined high central dark matter density
and the proximity of Segue\,1 make it an attractive target for
indirect dark matter detection experiments.

The number of ultra-faint dSphs around the Milky Way has doubled in
the past few years.  The seemingly ubiquitous presence of these
objects has forced a fundamental shift in galaxy formation models at
the smallest scales.  The high $M/L$ ratios and dark matter densities
of the least luminous dSphs, such as Segue\,1, may also lead to an
improved understanding of dark matter itself.  The promised discovery
of many additional ultra-faints dwarfs around the Milky Way and other
nearby galaxies makes this fertile ground for continued study.

\acknowledgments

We acknowledge and appreciate conversations regarding this work with
James Bullock, Raja Guhathakurta, Manoj Kaplinghat, Shane Walsh and
Adi Zolotov.  ENK acknowledges the support of an NSF Graduate Research
Fellowship.  JDS gratefully acknowledges the support of a Millikan
Fellowship provided by Caltech.  LES by NASA through Hubble Fellowship
grant HF-01225.01 awarded by the Space Telescope Science Institute,
which is operated by the Association of Universities for Research in
Astronomy, Inc., for NASA, under contract NAS 5-26555.






\clearpage

\begin{deluxetable}{lccr}
\tabletypesize{\scriptsize}
\tablecaption{Summary of Observed and Derived Quantities for Segue\,1}
\tablewidth{0pt}
\tablehead{
\colhead{Row} & \colhead{Quantity} & \colhead{Units} & \colhead{Segue\,1}
}
\startdata
(1) & RA        & h:m:s          & 10:07:03.2$\pm1.7^{s}$ \\
(2) & DEC       & $^{\circ}: \> ': \> ''$ & +16:04:25$\pm15''$ \\ 
(3) & E(B-V)    & mag            & 0.032 \\
(4) & Dist      & kpc            & $23\pm2$  \\
(5) & $M_{V,0}$  & mag            & $-1.5^{+0.6}_{-0.8}$ \\
(6) & $L_{V,0}$  & \Lsun          & 340\\
(7) & $\epsilon$&                & $0.48^{+0.10}_{-0.13}$ \\
(8) & $\mu_{V,0}$ & mag arcs$^{-2}$ & $27.6^{+1.0}_{-0.7}$ \\ 
(9)  & $r_{\rm eff}$ & $'$        & $4.4^{+1.2}_{-0.6}$  \\
(10)  & $r_{\rm eff}$ & pc          & $29^{+8}_{-5}$  \\
\hline
(11)  & $v$         & \kms        & $206.4\pm1.3$\\
(12)  & $v_{\rm GSR}$  & \kms        & $114\pm2$\\
(13)  & $\sigma$  & \kms           & $4.3\pm1.2$\\
(14)  & Mass     & \Msun           & $4.3^{+4.7}_{-2.5}\times10^5$  \\
(15)  & M/L      & \Msun/\Lsun     & $1340^{+4340}_{-990}$    \\
(16)  & [Fe/H]   & dex             & $-3.3\pm 0.2$  
\enddata
\tablecomments{Columns (1)-(2) and (5)-(10) taken from the SDSS
  photometric analysis of \citet{martin08a}.  Column (3) from
  \citet{schlegel98a} and (4) from \citet{belokurov06b}.  Columns
  (11)-(16) are derived in \S\,\ref{sec_kin}.}
\end{deluxetable}

\begin{deluxetable}{lccccccccc}
\tabletypesize{\scriptsize}
\tablecaption{Keck/DEIMOS Velocity Measurements for Stars in Segue\,1 Sample}
\tablewidth{0pt}
\tablehead{
\colhead{i} &
\colhead{Name} &
\colhead{$\alpha$ (J2000)} &
\colhead{$\delta$ (J2000)} &
\colhead{$g$} &
\colhead{$(g-r)$} &
\colhead{$v$} &
\colhead{$v_{\rm err}$} &
\colhead{$v_{\rm gsr}$} \\
\colhead{}&
\colhead{}&
\colhead{h$\,$ $\,$ m$\,$ $\,$s} &
\colhead{$^\circ\,$ $\,'\,$ $\,''$} &
\colhead{mag} &
\colhead{mag} &
\colhead{\kms} &
\colhead{\kms} &
\colhead{\kms} &
}
\startdata
\multicolumn{9}{c}{{\bf Segue\,1 Members}}\\
 1 &  3451635   &     10:06:40.5 &    +16:02:38.1 &     22.0 &    0.36 &    204.1 &      6.4 &    109.2 \\
 2 &  3451345   &     10:06:44.5 &    +16:01:29.4 &     20.7 &    0.27 &    210.5 &      4.0 &    115.5 \\
 3 &  3451159   &     10:06:44.6 &    +15:59:53.9 &     17.3 &   -0.01 &    200.4 &      2.2 &    105.5 \\
 4 &  3451358   &     10:06:49.1 &    +16:03:48.7 &     20.6 &    0.22 &    198.9 &      5.1 &    104.0 \\
 5 &  3451685   &     10:06:49.6 &    +16:03:08.3 &     21.1 &    0.13 &    207.8 &      6.7 &    112.9 \\
 6 &  3451364   &     10:06:52.3 &    +16:02:35.8 &     18.9 &    0.48 &    215.6 &      2.9 &    120.7 \\
 7 &  3451423   &     10:06:55.4 &    +16:04:16.2 &     20.7 &    0.27 &    213.0 &      3.8 &    118.1 \\
 8 &  3451533   &     10:06:57.4 &    +16:03:00.0 &     21.6 &    0.29 &    216.8 &      4.2 &    121.9 \\
 9 &  3451726   &     10:06:57.6 &    +16:02:30.1 &     22.3 &    0.14 &    212.8 &      5.3 &    117.9 \\
10 &  3451735   &     10:06:59.8 &    +16:02:18.5 &     22.0 &    0.38 &    203.6 &      4.9 &    108.7 \\
11 &  3451382   &     10:07:03.2 &    +16:03:35.0 &     21.8 &    0.34 &    206.6 &      5.2 &    111.7 \\
12 &  3451378   &     10:07:03.3 &    +16:02:34.4 &     20.6 &    0.29 &    205.5 &      2.7 &    110.6 \\
13 &  3451306   &     10:07:05.6 &    +16:04:22.0 &     17.5 &   -0.08 &    198.7 &      2.3 &    103.8 \\
14 &  3451374   &     10:07:01.3 &    +16:02:00.0 &     20.5 &    0.25 &    208.7 &      2.7 &    113.8 \\
15 &  3451757   &     10:07:01.5 &    +16:03:04.4 &     22.4 &    0.12 &    200.4 &      6.1 &    105.5 \\
16 &  3451790   &     10:07:06.7 &    +16:04:44.4 &     21.8 &    0.29 &    206.5 &      6.7 &    111.6 \\
17 &  1894468   &     10:07:14.8 &    +16:06:27.1 &     22.7 &    0.57 &    205.2 &      5.4 &    110.3 \\
18 &  3517005   &     10:07:14.9 &    +16:04:48.8 &     21.0 &    0.27 &    207.2 &      3.7 &    112.3 \\
19 &  1894643   &     10:07:15.1 &    +16:07:08.2 &     21.7 &    0.39 &    206.3 &      6.5 &    111.4 \\
20 &  3517002   &     10:07:15.7 &    +16:03:00.0 &     21.2 &    0.14 &    206.4 &     13.1 &    111.5 \\
21 &  3517007   &     10:07:16.3 &    +16:03:40.3 &     21.7 &    0.26 &    198.4 &      4.4 &    103.5 \\
22 &  3516925   &     10:07:24.1 &    +16:04:29.9 &     22.1 &    0.32 &    197.1 &      7.6 &    102.2 \\
23 &  1894761   &     10:07:28.4 &    +16:07:41.2 &     22.4 &    0.56 &    216.9 &     14.3 &    122.0 \\
24 &  3517048   &     10:07:31.1 &    +16:04:19.5 &     21.6 &    0.32 &    212.4 &      9.8 &    117.5 \\
\hline \\
\multicolumn{9}{c}{{\bf Higher Velocity Stars}}\\
1   &  3451696   &     10:06:50.8 &    +16:03:51.2 &     22.1 &    0.16 &    312.1 &     11.9 &    217.2 \\
2   &  3517146   &     10:07:13.7 &    +16:04:44.8 &     22.1 &    0.35 &    299.7 &      3.6 &    204.8 \\
3   &  3516836   &     10:07:17.4 &    +16:03:55.6 &     20.1 &    0.32 &    295.4 &      2.4 &    200.5 \\
4   &  3517243   &     10:07:32.5 &    +16:05:00.5 &     22.6 &    0.52 &    280.6 &      6.9 &    185.7 \\
\hline \\
\multicolumn{9}{c}{{\bf Non-Members}}\\
1   &  3451324   &     10:06:35.5 &    +16:02:21.1 &     17.7 &    0.88 &      1.4 &      2.2 &    $-$93.5 \\
2   &  3451835   &     10:06:36.3 &    +16:02:46.3 &     23.2 &    1.22 &    $-$21.2 &      2.7 &   $-$116.1 \\
..   &     ..     &         ..      &         ..      &       ..  &   ..     &       ..  &      ..  &     ..    
\enddata
\tablecomments{Velocity measurements for member stars of Segue\,1,
  higher velocity stars possibly associated with the Sgr stream and
  non-members.  Positions and magnitudes are taken from the SDSS DR6.
  We list the heliocentric radial velocity ($v$), velocity error
  ($v_{\rm err}$), and Galactocentric velocity ($v_{\rm gsr}$) for each
  star as determined in \S\,\ref{subsec_rvel}.  Entries for
  non-members are published in their entirety in the electronic
  edition of the {\it Astrophysical Journal}.  }
\end{deluxetable}

\clearpage


\end{document}